\documentclass[twoside,reqno]{HERON}
\usepackage{epsfig,cite,colordvi}
\usepackage{graphicx}
\usepackage{url}
\usepackage{amssymb,amsmath,amscd,epsf}
\usepackage{times}
\usepackage{makeidx}
\begin{document}

\title{Microscopic Optical Potential for Elastic Proton-Nucleus Scattering from Chiral Forces}

\runningheads{Microscopic Optical Potential for Elastic Proton-Nucleus Scattering from Chiral Forces}{C. Giusti, M. Vorabbi, P. Finelli}

\begin{start}

\author{C. Giusti}{1}, \coauthor{M. Vorabbi}{2}, \coauthor{P. Finelli}{3}

\index{Giusti, C.}
\index{Vorabbi, M.}
\index{Finelli, P.}

\address{Dipartimento di Fisica,  
Universit\`a degli Studi di Pavia and
INFN, Sezione di Pavia,  Via A. Bassi 6, I-27100 Pavia, Italy}{1},

\address{TRIUMF, 4004 Wesbrook Mall, Vancouver, British Columbia, V6T 2A3 Canada}{2},

\address{Dipartimento di Fisica e Astronomia, Universit\`{a} degli Studi di Bologna and INFN, Sezione di Bologna, Via Irnerio 46, I-40126 Bologna, Italy}{3},

\begin{Abstract}

A microscopic optical potential (OP) is derived from $NN$ chiral potentials 
at the first-order term within the spectator expansion of the multiple scattering theory and adopting the impulse approximation. The performances of our OP are compared with those of a phenomenological OP in the description of elastic proton scattering data on different isotopic chains. An analogous scheme is adopted to construct a microscopic OP for elastic antiproton-nucleus scattering. The results of our OPs are in reasonably good agreement with the experimental data, for both elastic proton and antiproton-nucleus scattering.

\end{Abstract}
\end{start}

\section{Introduction}
The optical potential (OP) provides a suitable tool to describe elastic nucleon-nucleus ($NA$) scattering. Its use can be extended to inelastic scattering and to perform calculations for a wide variety of nuclear reactions. 
In usual calculations phenomenological OPs are adopted, that are obtained assuming an analytical form and a dependence on a number of adjustable parameters for the real and imaginary parts (the OP is complex) that characterize the shape of the nuclear density distribution and that vary with the nucleon energy and the nuclear mass number (the OP is energy dependent and can depend on the nuclear mass number $A$). The values of the parameters are determined through a fit to elastic $pA$ scattering data. 
Alternatively and more fundamentally, the OP can be obtained from a microscopic calculation, which, in principle, requires the solution of the full many-body nuclear problem for the incident nucleon and the $A$ nucleons of the target and therefore represents a very hard and challenging task.
Several approximations are required to reduce the complexity of the original problem.
In general, we do not expect that a theoretical OP, which is the result of several approximations, will be able to describe elastic $NA$ scattering data better than a phenomenological OP whose parameters have been fitted to data, in particular, if we consider data included in the database used for the fitting procedure, but it might have a greater predictive power in situations for which experimental data are not yet available. 

In Refs.~\cite{Vorabbi1,Vorabbi2,Vorabbi3} we derived a microscopic OP for elastic $pA$ scattering from $NN$ chiral potentials up to fourth (N$^3$LO) and fifth (N$^4$LO) order in the chiral perturbative expansion. Recently, we have derived, within an analogous scheme, a microscopic OP for elastic $\bar{p}A$ scattering~\cite{Vorabbi4}. Our first purposes were to study the domain of applicability of microscopic two-body chiral potentials, to check the convergence, and to assess the theoretical errors associated with the truncation of the chiral expansion in the construction of an OP.

Our OP has been obtained within a theoretical framework based on the Watson multiple scattering theory~\cite{Watson}. We adopted several approximations, with the idea to start from a relatively simple model that can then be improved.  

Our contribution is organized as follows: In Section~2 we outline the theoretical framework used to calculate our microscopic OP.  In
Section~3 we discuss its performances in comparison with  elastic $pA$ scattering data on different nuclei and isotopic chains. Our results are compared with those of the successful phenomenological OP of Ref.~\cite{KD,KON}. In Section~5 we present our OP for elastic $\bar{p}A$ scattering. Some conclusions are drawn in Section~5.

\section{Theoretical framework}
Proton elastic scattering off a target nucleus with $A$ nucleons can be formulated in the 
momentum space by the full Lippmann-Schwinger (LS) equation 
\begin{equation}
\label{generalscatteq}
T = V \left( 1 + G_0 (E) T \right)\, , 
\end{equation}
where $V$ represents the external interaction which, if we assume only two-body forces, is given by the sum over all the target nucleons  of two-body potentials describing the interaction of each target nucleon with the incident proton, and $G_0 (E)$ is the free Green's function for the $(A+1)$-nucleon system.

As a standard procedure, Eq.~(\ref{generalscatteq}) is separated into a set of two coupled integral equations: the first one for the $T$ matrix
\begin{equation}\label{firsttamp}
T = U \left(1+ G_0 (E) P T\right) \, 
\end{equation}
and the second one for the OP $U$
\begin{equation}\label{optpoteq}
U = V \left(1+ G_0 (E) Q U \right)\, .
\end{equation}

A consistent framework to compute $U$ and $T$ is provided by the spectator expansion, that is based on the multiple scattering theory~\cite{Watson}. We retain only the first-order term, corresponding to the single-scattering approximation, where only one target-nucleon interacts with the projectile. In addition, we adopt the impulse approximation, where nuclear binding forces on the interacting target nucleon are neglected~\cite{Vorabbi1}.

After some manipulations, the OP is obtained in the so-called optimum factorization approximation as the product of the free
$NN$ $t$ matrix and the nuclear matter densities
\begin{equation}
\label{optimumfact}
U ({\bf q},{\bf K};\omega) = \frac{A-1}{A} \, \eta ({\bf q},{\bf K}) 
 \sum_{N = n,p} t_{pN} \left({\bf q}, {\bf K}, \omega \right) \, \rho_N (q) \, ,
\end{equation}
where ${\bf q}$ and ${\bf K}$ are the momentum transfer and the total 
momentum, respectively, in the $NA$ reference frame, 
$t_{pN}$ represents the proton-proton ({\it pp}) and proton-neutron ({\it pn}) 
$t$ matrix, $\rho_N$ represents the neutron and proton profile density,
and $\eta ({\bf q},{\bf K})$ is the M\o ller factor, that imposes the 
Lorentz invariance of the flux when we pass from the $NA$ to the $NN$
frame in which the $t$ matrices are evaluated.  
Through the dependence of $\eta$ and $t_{pN}$ upon ${\bf K}$, the factorized OP in Eq.~(\ref{optimumfact}) exhibits nonlocality and off-shell effects~\cite{Vorabbi1}. 

\section{Results for elastic proton-nucleus scattering}

Two basic ingredients are required to calculate the OP in 
Eq.~(\ref{optimumfact}):
the $NN$ potential and the neutron and proton densities of the target nucleus.
For the densities we use a relativistic mean-field (RMF)
description~\cite{Nik1}, which has been quite successful in the description of ground state and excited state properties of finite nuclei, in particular in a density dependent meson exchange (DDME) version~\cite{Nik2}.
For the $NN$ interaction we have used in Ref.~\cite{Vorabbi1} two different versions of chiral potentials at fourth order (N$^3$LO) in the chiral expansion, presented by Entem and Machleidt (EM)~\cite{EM} and Epelbaum, Gl\"ockle, and Mei\ss ner (EGM)~\cite{EGM}, and in Ref.~\cite{Vorabbi2} the more recent $NN$ potentials at fifth order (N$^4$LO), presented by Epelbaum, Krebs, and Mei\ss ner (EKM)~\cite{EKM} and Entem, Machldeidt, and Nosyk (EMN)~\cite{EMN}.

The two versions of chiral potentials at N$^3$LO use different regularization prescriptions to treat divergent terms. In general, the integral in the LS equation is divergent and needs to be regularized. A usual procedure is to multiply the $NN$ potential entering the LS equation by a regulator function $f^\Lambda$. Both EM and EGM present results with three values of the cutoff parameter $\Lambda$ (450, 500, 600 MeV for EM and 450, 550, and 600 MeV for EGM), and treat differently the short-range part of the two-pion exchange contribution, that has unphysically strong attraction: EM adopt a dimensional regularization and EGM a spectral function regularization which introduces an additional cutoff $\tilde{\Lambda}$ and    
give the following cutoff combinations: $\{\Lambda, \tilde{\Lambda}\} =$ $\{450,500\}$, $\{450,700\}$, $\{550,600\}$, $\{600,600\}$, $\{600,700\}$. The sensitivity to the choice of the cutoff parameters and the 
order-by-order convergence of the chiral perturbation theory (ChPT) expansion have been investigated comparing the results produced by the different chiral potentials with available experimental data for the $NN$ scattering amplitudes and for the observables  (differential cross section $d\sigma/d\Omega$, 
analyzing power $A_y$, and spin rotation $Q$) of elastic proton scattering off ${}^{16}$O~\cite{Vorabbi1}.

Concerning the convergence, the results show that it is mandatory to use chiral potentials at N$^3$LO: potentials at lower orders produce results in clear disagreement with the experimental $NN$ scattering amplitudes
and with the observables of elastic $pA$ scattering~\cite{Vorabbi1}.
All the potentials at N$^3$LO reproduce the experimental amplitudes at $100$ MeV. The agreement becomes, as expected, worse upon increasing the energy and at $200$ MeV the set of potentials with lower cutoffs fail to reproduce empirical data~\cite{Vorabbi1}. 

\begin{figure}[ht]
\begin{minipage}[h]{0.5\textwidth}
\includegraphics[width=5.5cm]{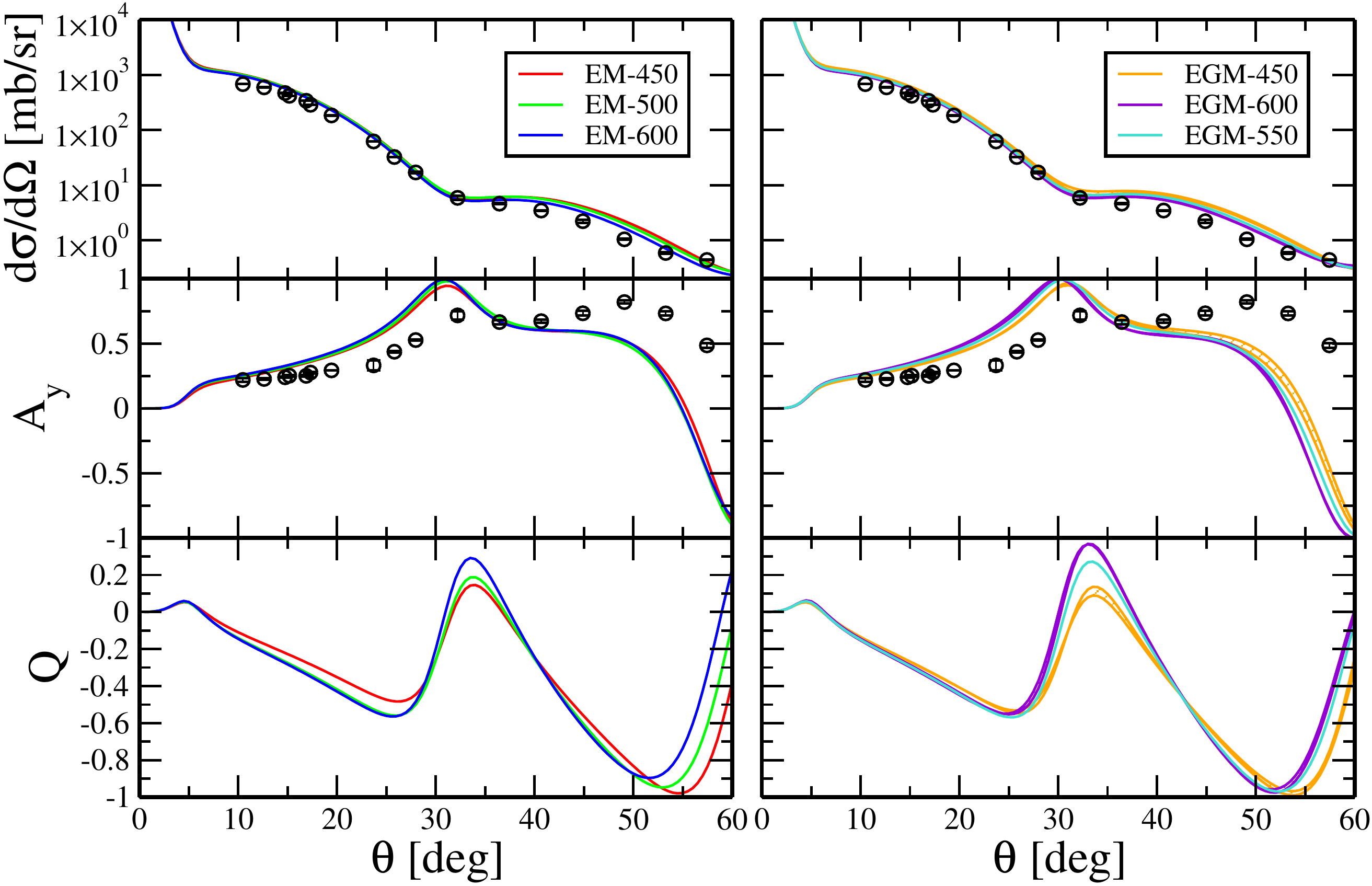}
\end{minipage} \hspace{0.01\textwidth}
\begin{minipage}[h]{0.5\textwidth}
\includegraphics[width=5.5cm]{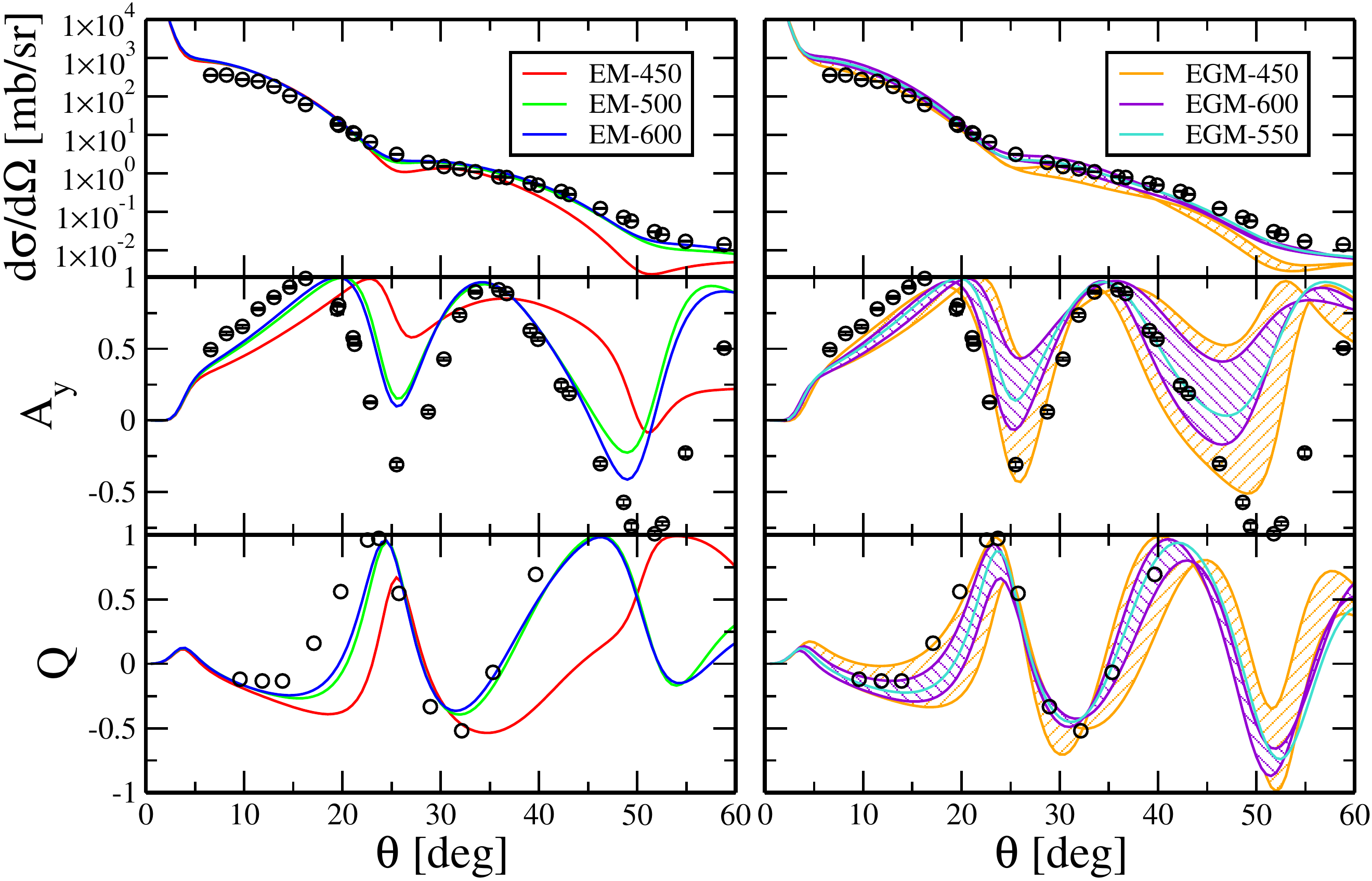}
\end{minipage}
\caption{Scattering observables  as a function of the center-of-mass scattering angle $\theta$  
for elastic proton scattering on ${}^{16}$O  at a laboratory energy of $100$ MeV (left figure) and $200$ MeV (right figure). The results obtained with the EM~\cite{EM} and EGM~\cite{EGM}  $NN$ chiral potentials at N$^3$LO are denoted by the value of the LS cutoff.
Data are taken from~\cite{kelly,exfor}.}
\label{fig:16O}
\end{figure}

In Figure~\ref{fig:16O} the observables for elastic proton scattering off ${}^{16}$O computed at $100$ MeV and $200$ MeV with the different $NN$ potentials at N$^3$LO  are displayed and compared with the empirical data. 
All sets of potentials give close results, with the exception of $A_y$ above $50$ degrees, where
all potentials overestimate the data up to the maximum and then display an unrealistic downward trend, and $Q$ around the maximum at $30$ degrees. 
In particular, the experimental cross section is well reproduced by all potentials in the minimum region, between $30$ and $35$ degrees. Polarization observables, which are more sensitive to the differences in the potentials and to the approximations of the model, are usually more difficult to reproduce.
At $200$ MeV EM and EGM with the lower cutoffs ($\Lambda = 450$ MeV) give results in clear disagreement with the empirical data, which are well described by the potentials with higher cutoffs.

The results with the  microscopic OP derived from the $NN$ potentials at N$^4$LO EKM~\cite{EKM} and EMN~\cite{EMN} indicate that the order-by-order convergence pattern is clear and that robust convergence has been reached at N$^4$LO. We do not expect large contributions from the higher-order extension in the $NN$ sector. The agreement of the theoretical results with the data is comparable, neither better nor worse, than the agreement obtained with chiral potentials at  N$^3$LO. A better agreement would require a better model for the OP,  where the approximations adopted in the present calculations of the OP are reduced.  


Although obtained assuming several approximations, our microscopic OP does not contain phenomenological inputs. In contrast, phenomenological OPs are based on the use of some free parameters, specifying the well and the geometry of the system, that are determined by a fitting procedure over a set of available data of elastic $pA$ scattering. The phenomenological approach provides OPs able to give an excellent description of data in many regions of the nuclear chart and for energy ranges where data are available, but which may lack predictive power when applied to situations where data are not yet available. We have seen that our OP gives a reasonable description of elastic $pA$ scattering data without the need to introduce parameters fitted to empirical data.  Being the result of a model and not of a fitting procedure, a microscopic OP might have a more general predictive power than a phenomenological OP, but the approximations adopted to reduce the complexity of the original many-body problem might give a poorer agreement with available data. In order to investigate and clarify this issue, it can be useful to compare the performances of our microscopic OP and of a successful phenomenological OP in the description of elastic proton scattering data on nuclei of some isotopic chains. For the comparison we have considered the phenomenological OP of Refs.~\cite{KD,KON} (KD). A systematic investigation has been performed in a range of  proton energies around and above 200 MeV~\cite{Vorabbi3}, with the aim to test the upper energy limit of applicability of our OP before the chiral expansion scheme breaks down.

The nonrelativistic phenomenological KD potential~\cite{KD} is a so-called "global" OP, which means that the free adjustable parameters are fitted for
a wide range of nuclei ($24 \le A \le 249$) and of incident energies ($1$ keV $\le E \le 200$ MeV) with some parametric dependence of the coefficients in terms of $A$ and $E$. 
Recently, an extension of KD up to 1 GeV has been proposed~\cite{KON}, with the aim to test at which energy the
predictions of a nonrelativistic phenomenological OP fail. 
Above 200 MeV an approach based on the Dirac equation would probably be a more consistent choice, but, since we are interested in testing the limit of applicability of our (nonrelativistic) microscopic OP,  we have employed such an extension for our present purposes. 
All the calculations have been performed by ECIS-06 \cite{ecis} as a subroutine in the TALYS software \cite{KON,talys}.

The microscopic OP adopted for the comparison has been derived from  the two $NN$ chiral potentials at N$^4$LO, EKM~\cite{EKM} and EMN~\cite{EMN}, which differ in the renormalization procedures. The strategy followed for the EKM potentials~\cite{EKM} consists in a coordinate space regularization for the long-range contributions $V_{\rm long} ({\bf r}) $, by the introduction of $f\left(\frac{r}{R} \right) = \left(1 -\exp\left( -\frac{r^2}{R^2}\right) \right)^n$, and a conventional momentum space regularization for the contact  (short-range) terms, with a cutoff $\Lambda = 2R^{-1}$. Five choices of $R$ are available ($0.8, 0.9, 1.0, 1.1$, and $1.2$ fm).
For the EMN potentials~\cite{EMN} a spectral function regularization, with a cutoff $\tilde{\Lambda} \simeq 700$ MeV, was employed to regularize the loop contributions and a conventional regulator function,
with $\Lambda = 450,500$, and $550$ MeV, to deal with divergences in the LS equation. 

\begin{figure}[ht]
\begin{center}
\includegraphics[width=9cm]{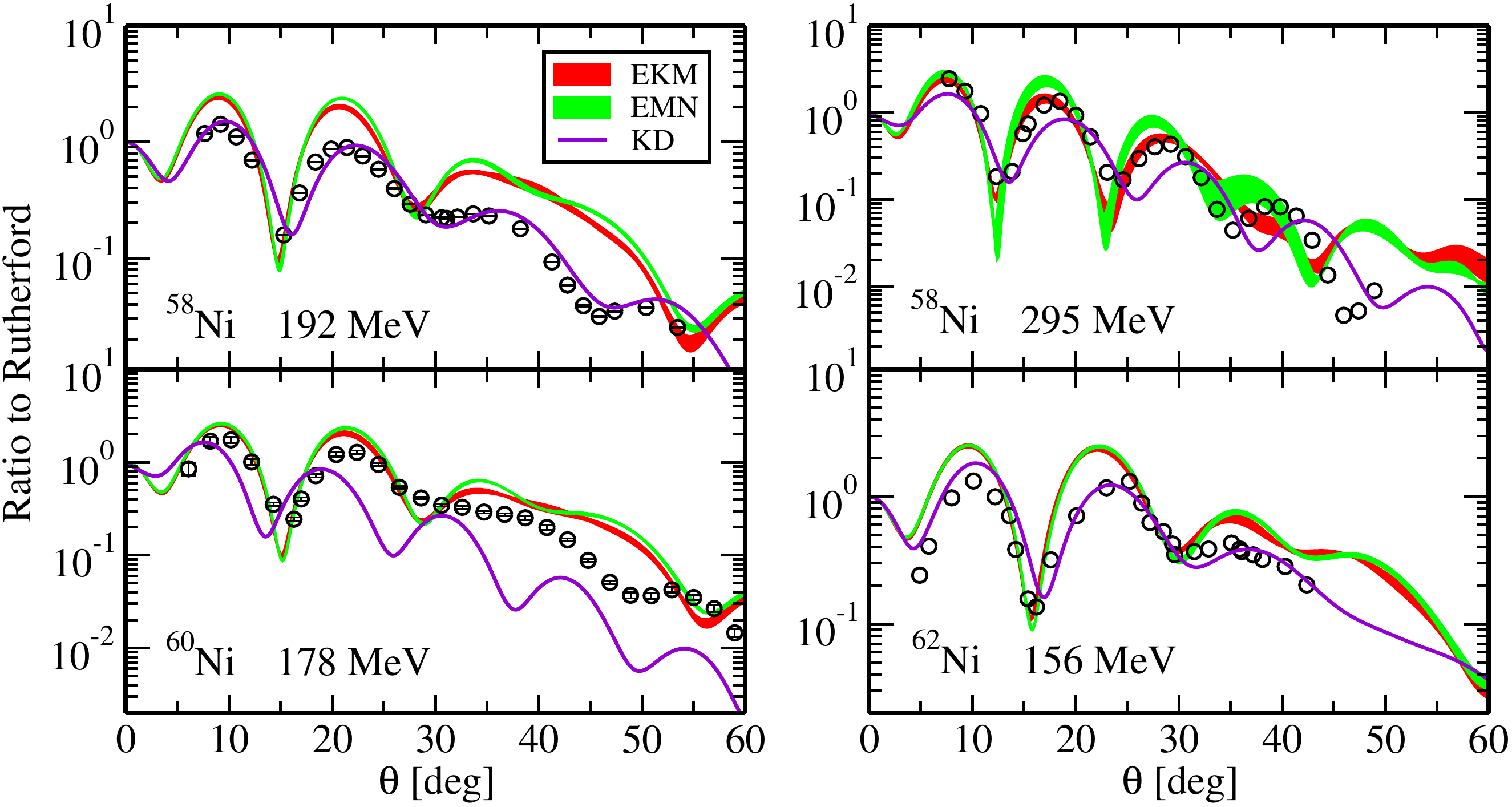}
\caption{Ratio of the differential cross section to the Rutherford cross section as a function of 
the center-of-mass scattering angle $\theta$ for elastic proton scattering on Ni isotopes: 
${}^{58}$Ni at $E = 192$ and $295$ MeV,  ${}^{60}$Ni at $E = 178$ MeV, and  ${}^{62}$Ni 
at $E = 156$ MeV. In the calculations the microscopic OPs derived from the
EKM~\cite{EKM} (red band)  and EMN~\cite{EMN} (green band) NN chiral potentials at N$^4$LO and with the phenomenological global OP of Ref.~\cite{KON} 
(KD, violet line).  Experimental data from Ref. \cite{kelly,exfor}. }
\label{fig:ni}
\end{center}
\end{figure}

\begin{figure}[ht]
\begin{center}
\includegraphics[width=9cm]{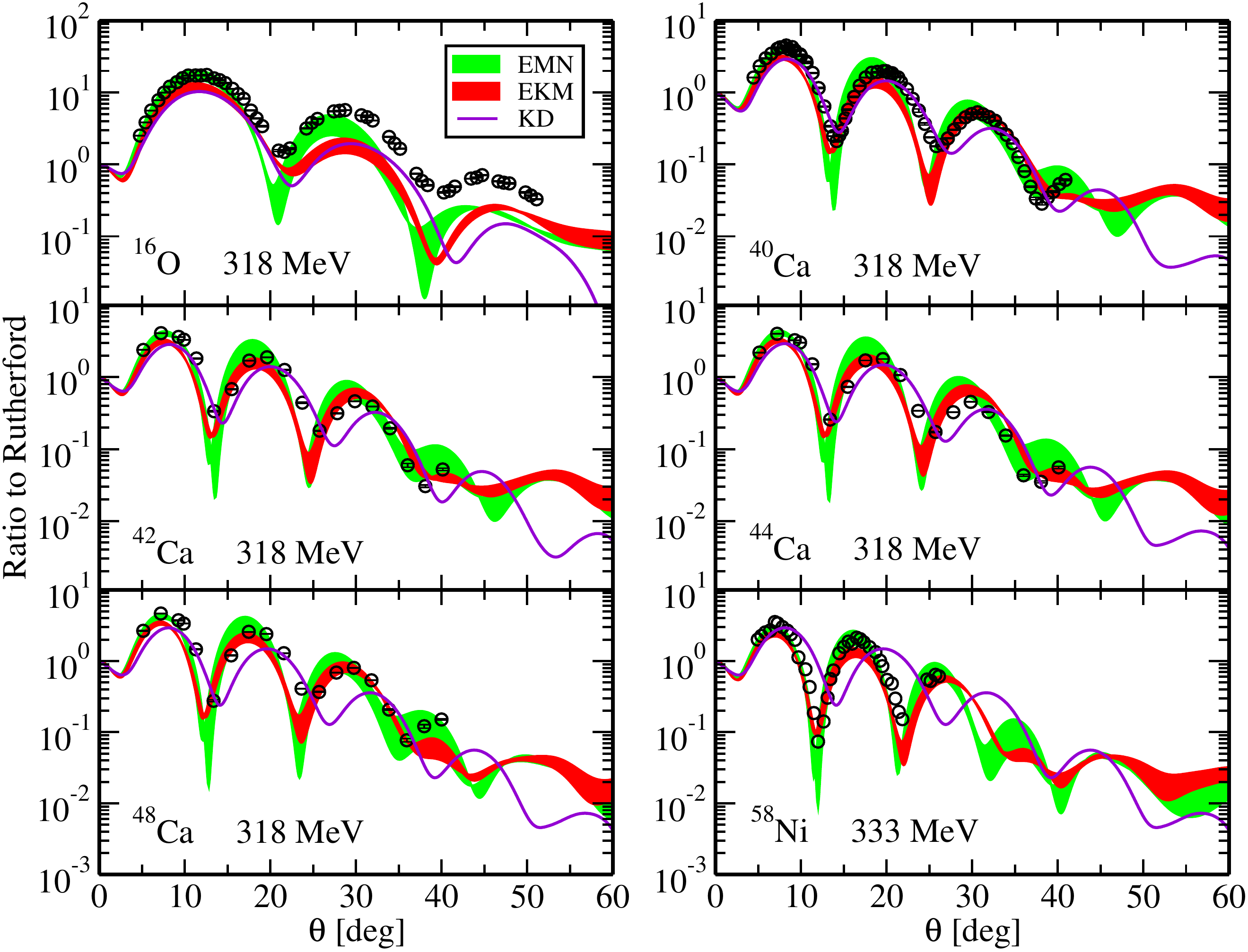}
\caption{The same as in Fig.~\ref{fig:ni} for ${}^{16}$O and ${}^{40,42,44,48}$Ca 
at $E = 318$ MeV and ${}^{58}$Ni at $E = 333$ MeV.  Experimental data from Ref. \cite{kelly,exfor}. }
\label{fig:high}
\end{center}
\end{figure}

If we want to test the predictive power of our OP in comparison with available data it can be useful to show the uncertainties produced by different values of the regularization parameters.
For this purpose, we have performed calculations  with $R=0.8, 0.9$, and $1.0$ fm for EKM and with $\Lambda = 500$  and $550$ MeV for EMN. The bands in Figure~\ref{fig:ni} and \ref{fig:high} give the differences produced by changing $R$ for EKM (red bands) and $\Lambda$ for EMN (green bands).

Calculations have been performed for proton energies between 156 and 333 MeV. The energy range was chosen on the basis of the approximations adopted to derive our OP, in particular, the impulse approximation does not allow us to use our OP with enough confidence at much lower energies. The upper energy limit is determined by the fact that EKM and EMN are able to describe $NN$ scattering observables up to 300 MeV \cite{EKM,EMN}. 

The ratios of the differential cross sections to the Rutherford cross sections for elastic proton scattering off nichel isotopes are shown in Figure~\ref{fig:ni}.
Data for ${}^{58}$Ni up to 200 MeV and ${}^{60}$Ni up to 65 MeV are included in the experimental database used to generate the KD potential, which  gives an excellent description of  ${}^{58}$Ni data  at 192 MeV but a much worse agreement at 295 MeV, where it is able to describe only the overall behavior of the experimental cross section. 
The EKM and EMN results provide a better and reasonable description of the data at 295 MeV, up to $\theta \sim 40^{\circ}$, while at 192 MeV they give a rough description of the shape of the experimental cross section but the size is somewhat overestimated. KD gives only a poor description of the data for ${}^{60}$Ni at 178 MeV and a very good agreement 
for ${}^{62}$Ni at 156 MeV. The microscopic OP gives a better  agreement with  the ${}^{60}$Ni data, 
while for ${}^{62}$Ni the results are a bit larger than those of the KD potential. 
The EKM and EMN results are always very close to each other and the bands, representing the theoretical uncertainties produced by  
different values of the regularization parameters, are generally narrow.

The results for different isotopic chains~\cite{Vorabbi2} indicate that our microscopic OP has a comparable and in some cases even better predictive power than the KD potential in the description of the experimental cross sections. KD gives a better and excellent description of data, in particular, of data included in the database used to generate the KD potential and at the lower energies considered. Above 200 MeV our OP gives, in general, a better agreement with the data. This conclusion is confirmed in Figure~\ref{fig:high}, where numerical and experimental results are compared for elastic 
proton scattering off ${}^{16}$O and ${}^{40,42,44,48}$Ca at $E = 318$ MeV and ${}^{58}$Ni at $E = 333$ MeV. The differences between the phenomenological 
and microscopic OPs increase with increasing scattering angle and proton energy. For ${}^{58}$Ni at 333 MeV 
both EKM and EMN give a much better and good description of the data. In the other cases KD is able to describe the data only at the lowest angles. 

\section{Elastic Antiproton-Nucleus Scattering}
With the advent of new facilities, namely the Extreme Low ENergy Antiproton (ELENA) ring at CERN~\cite{puma} and the Facility for Antiproton and Ion Research (FAIR)~\cite{fair} in Germany, scientific interest in new experiments on antiproton scattering off nuclear targets will experience a renaissance. 

The dominant feature of $\bar{p}p$ scattering at low energies is the annihilation process that, due to its large cross-section, greatly reduces the probability of rescattering processes. Therefore $\bar{p}A$ scattering is likely to be described 
without the complication of multiple scattering processes, which makes it a  {\it clean} method to study nuclear properties. 
The $\bar{p}$ absorption is surface-dominated~\cite{walcher,dover,adachi} and is sensitive to nuclear radii. 
The exchange mechanism and the antisymmetrization between the projectile and the target constituents are not relevant in the $\bar{p}A$ interaction, while
the role played by the three-body $\bar{p}NN$ forces still remains an open question.

We have derived a microscopic OP for elastic $\bar{p}A$
scattering~\cite{Vorabbi4} following a scheme analogous to that employed for elastic $pA$ scattering~\cite{Vorabbi1,Vorabbi2,Vorabbi3,gennari} and using the most recent techniques in nuclear physics. 
The density matrix has been obtained using the same approach as in Ref.~\cite{gennari}, where one-body translationally invariant (trinv) nonlocal densities were computed within the {\it ab initio} No-Core Shell Model~\cite{barrett} (NCSM) approach using two- and three-nucleon chiral interactions as the only input.
We used the $NN$ chiral interaction of Ref.~\cite{EMN,Entem} up to N$^4$LO
and the $NNN$ chiral interaction up to N$^2$LO, which employs a simultaneous local and nonlocal regularization with the cutoff values of
$650$ MeV and $500$ MeV, respectively~\cite{Navratil2007,gysbers}.  
Details 
can be found in Ref.~\cite{gennari}.
We note that the use of a nonlocal density requires an unfactorized OP~\cite{gennari}. 

The same $NN$ interaction used for the calculation of the nuclear density was used in Ref.~\cite{gennari} to compute the $pA$ scattering matrix. The $\bar{p}N$ and  $pN$ interactions are different and in the case of the OP for  $\bar{p}A$ scattering it is not possible to compute the nuclear density and the  $t_{\bar{p}N}$ matrix with the same potential. We have used the first recently derived $\bar{p}N$ interaction at NN$^3$LO~\cite{Dai2017}. 

The main difference between $NN$ and $\bar{N}N$ is that in the $\bar{N}N$ case the annihilation channel is available because the total baryon number is zero. For low momentum protons, elastic $\bar{p}N$ requires a higher number of partial waves than the $pN$ counterpart. All phase shifts are complex because of the annihilation process and both isospin 0 and 1 contribute in each partial wave~\cite{PhysRevC.43.1610}.
As a consequence, a treatment of $\bar{p}N$ scattering is intrinsically more complex than the $NN$ system.

\begin{figure*}[t]
\begin{center}
\includegraphics[scale=0.3]{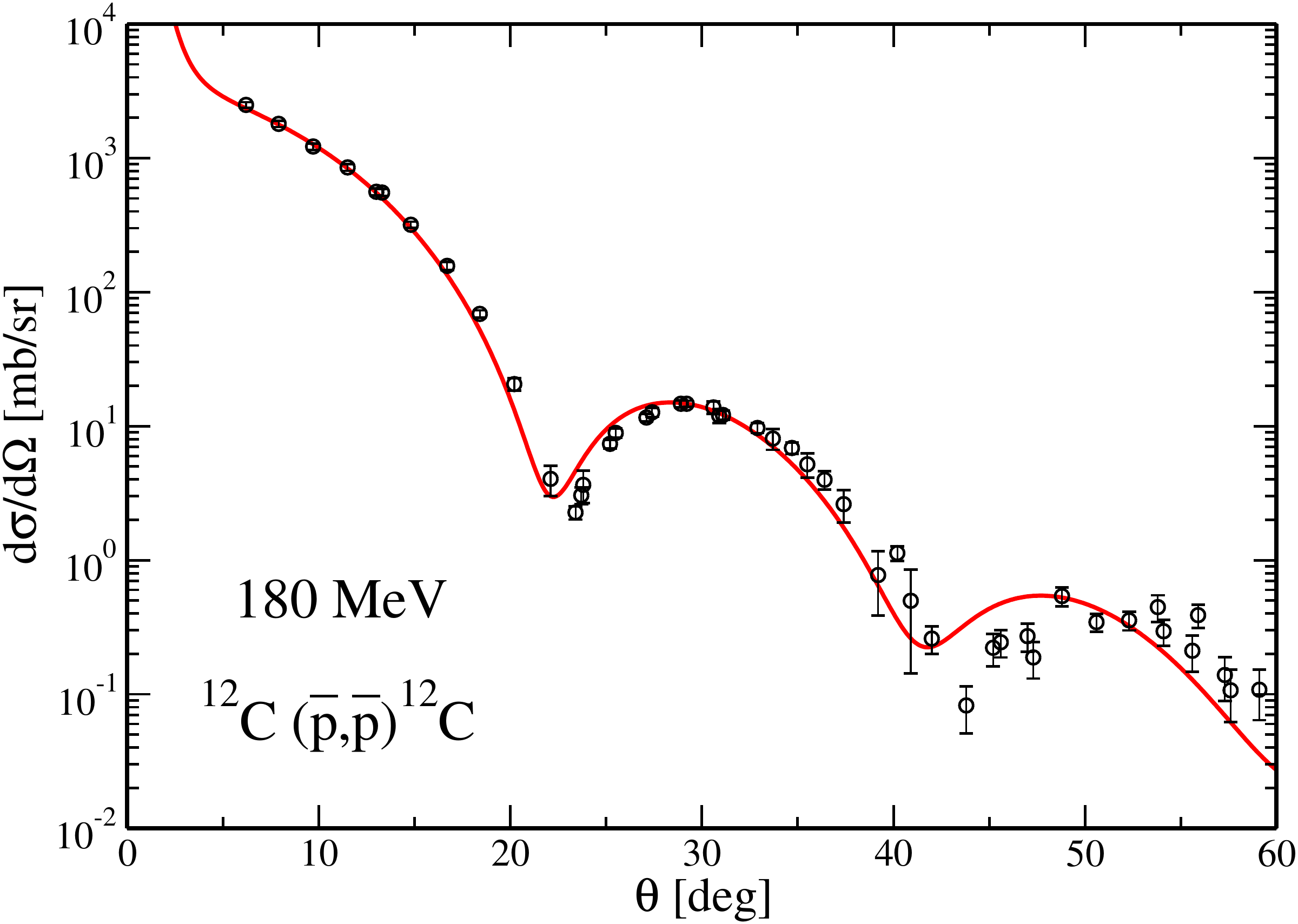}
\caption{Differential cross section as a function of the center-of-mass scattering angle for elastic antiproton scattering off $^{12}$C at the laboratory energy of 180 MeV. Experimental data from Refs.~\cite{batu,garreta,BRUGE198614}.
\label{fig_cross_sections} }
\end{center}
\end{figure*}
An example of our results is presented in Fig.~\ref{fig_cross_sections}, where the differential cross section of elastic antiproton scattering off $^{12}$C, computed at the laboratory energy of 180 MeV, is compared with the experimental data. Our microscopic OP describes the data very well. In particular, it is remarkable the agreement in correspondence of the first minimum of the diffraction pattern. More results for different target nuclei~\cite{Vorabbi4} confirm the ability of our microscopic OP to describe the empirical data  very well.

\section{Conclusions}
A microscopic OP for elastic $pA$ scattering has been derived as the first-order term within the spectator expansion of the nonrelativistic multiple scattering theory, adopting the impulse approximation, and neglecting medium effects. The calculation requires two basic ingredients: the nuclear density, which has been obtained within a relativistic mean-field description,  and the nuclear interaction. 
for which we have used different versions of $NN$ chiral potentials at N$^3$LO and N$^4$LO, which differ in the regularization prescriptions to treat divergent terms. 

Our first aims were to study the domain of applicability of two-body chiral potentials in the construction of an OP, to check the convergence assessing theoretical errors associated with the truncation of the ChPT expansion, and to compare the results produced by the different $NN$ chiral potentials on elastic $pA$ scattering observables.

Our work shows that building an OP within the ChPT is a promising approach for describing elastic $pA$ scattering. The  convergence pattern is clear and robust convergence has been reached at N$^4$LO. 

The performances of our OP have been compared with those of the phenomenological KD OP in the description of elastic proton scattering data on some isotopic chains. The agreement of our results with data is comparable with the predictions of  the KD potential, in particular for energies above 200 MeV. 

Following an analogous scheme, we have derived the OP for elastic $\bar{p}A$ scattering. In the calculations one-body translationally invariant nonlocal densities were computed within the {\it ab initio} No-Core Shell Model approach using two- and three-nucleon chiral interactions.  
The new $\bar{N}N$ interaction up to $\mathrm{N}^3\mathrm{LO}$ has been used to obtain the $t_{\bar{p}N}$ scattering matrix. Our result are in good agreement with the antiproton elastic scattering data. 

Although in many cases able to describe the experimental data, our OP contains several approximations and it can be improved. As possible improvements on which we plan to work in the near future we mention: 1) The inclusion of three-body forces in the nuclear potential for the scattering matrix; 2) To go beyond the impulse approximation and include nuclear medium effects.


\begin{thebibliography}{99}


\bibitem{Vorabbi1}
M. Vorabbi, P. Finelli, and C. Giusti, \textit{Phys. Rev.}, \textbf{C93} (2016) 034619.

\bibitem{Vorabbi2}
M. Vorabbi, P. Finelli, and C. Giusti, \textit{Phys. Rev.}, \textbf{C96} (2017) 044001.

\bibitem{Vorabbi3}
M. Vorabbi, P. Finelli, and C. Giusti, \textit{Phys. Rev.}, \textbf{C98} (2018) 064602.

\bibitem {Vorabbi4}
M.Vorabbi \textit{et al.}, arXiv:1906.11984.

\bibitem{Watson}
K.M. Watson, \textit{Phys. Rev.}, \textbf{89} (1953) 575-587.


\bibitem{KD}
A. J. Koning and J. P. Delaroche, \textit{Nucl. Phys.} \textbf{A713} (2003) 231-310.

\bibitem{KON}
A. J. Koning, S. Hilaire, and M.C. Duijvestijn, \textit{Proceedings of the International Conference on Nuclear Data for Science and Technology, April 22-27, 2007, Nice, France} (2008) 211-214.

\bibitem{Nik1}
T. Nik$\mathrm{\check{s}}$i$\acute{\mathrm{c}}$ \textit{et al.},  \emph{Computer Physics Communications} \textbf{185} (2014) 1808-1821.

\bibitem{Nik2}
T. Nik$\mathrm{\check{s}}$i$\acute{\mathrm{c}}$ \textit{et al.}, \emph{Phys. Rev.}, \textbf{C66} (2002) 024306.

\bibitem{EM}
D. R. Entem and R. Machleidt, \textit{Phys. Rev.} \textbf{C68} (2003) 041001.

\bibitem{EGM}
E. Epelbaum, W. Gl\"ockle, and U.-G. Mei\ss ner, \textit{Nucl. Phys.} \textbf{A747} (2005) 362-424.

\bibitem{EKM}
E. Epelbaum, H. Krebs, and U.-G. Mei\ss ner \textit{Eur. Phys. J} \textbf{A51} (2015) 53; \textit{Phys. Rev. Lett.} \textbf{115} (2015) 122301.

\bibitem{EMN}
D. R. Entem, R. Machleidt, and Y. Nosyk, \textit{Phys. Rev.} \textbf{C96} (2017) 024004.

\bibitem{kelly}
\textbf{http://www.physics.umd.edu/enp/jjkelly/datatables.htm}.

\bibitem{exfor}
\textbf{http://www.nndc.bnl.gov/exfor/exfor.htm}.

\bibitem{ecis}
J. Raynal, ``Notes on ecis94'' (1994).

\bibitem{talys}
\textbf{www.talys.eu/fileadmin/talys/user/docs/talys1.8.pdf}.

\bibitem{puma}
A. Obertelli \textit{et al.}, \textit{Letter of Intent} CERN-SPSC-I-247 (2017) 

\bibitem{fair}
\textbf{https://www.gsi.de/en/researchaccelerators/fair.htm}.

\bibitem{walcher}
T. Walcher, \textit{Annual Review of Nuclear and Particle Science}, \textbf{38} (1988) 67-95.

\bibitem{dover}
C. Dover, \textit{et al.}, \textit{Progress in Particle and  Nuclear Physics}, \textbf{29} (1992) 87-173.

\bibitem{adachi}
S. Adachi and H.V. von Geramb, \textit{Nucl. Phys.}, \textbf{A470} (1987) 461-476.

\bibitem{gennari}
M. Gennari, M. Vorabbi, A. Calci, and P. Navr{\'a}til, \emph{Phys. Rev.}, \textbf{C97} (2018) 034619.

\bibitem{barrett}
B. R. Barrett, P. Navr{\'a}til, and J. P. Vary \textit{Progress in Particle and  Nuclear Physics}, \textbf{69} (2013) 131-181.

\bibitem{Entem}
D. R. Entem \textit{et al.} \textit{Phys. Rev.} \textbf{C91} (2015) 014002.

\bibitem{Navratil2007}
P. Navr\'{a}til, \textit{Few-Body Syst} \textbf{41} (2007) 117-140.

\bibitem{gysbers}
P. Gysbers \textit{et al.} \textit{Nature Physics} \textbf{15} (2019) 428-431.

\bibitem{Dai2017}
L.-Y. Dai, J. Haidenbauer, and U.-G. Mei{\ss}ner,  \textit{JHEP} \textbf{07} (2017) 78.

\bibitem{PhysRevC.43.1610}
P. Bydzovsky, R. Mach, and F. Nichitiu,  \textit{Phys. Rev.} \textbf{C43} (1991) 1610-1618.


\bibitem{batu}
Yu. A. Batusov \textit{et al.},  \textit{Sov. J. Nucl. Phys.} \textbf{52} (1990) 776-781.

\bibitem{garreta}
D. Garreta \textit{et al.},  \textit{Phys. Lett.} \textbf{B149} (1984) 64-68.

\bibitem{BRUGE198614}
D. Brugel \textit{et al.},  \textit{Phys. Lett.} \textbf{B649} (1986) 14-16.


\end{thebibliography}
\end{document}